\documentclass[sigconf]{acmart}
\usepackage{breakurl}

\AtBeginDocument{%
  \providecommand\BibTeX{{%
    \normalfont B\kern-0.5em{\scshape i\kern-0.25em b}\kern-0.8em\TeX}}}

\copyrightyear{2022} 
\acmYear{2022} 
\setcopyright{acmlicensed}\acmConference[EASE 2022]{The International Conference on Evaluation and Assessment in Software Engineering 2022}{June 13--15, 2022}{Gothenburg, Sweden}
\acmBooktitle{The International Conference on Evaluation and Assessment in Software Engineering 2022 (EASE 2022), June 13--15, 2022, Gothenburg, Sweden}
\acmPrice{15.00}
\acmDOI{10.1145/3530019.3535344}
\acmISBN{978-1-4503-9613-4/22/06}

\begin{document}

\title{The Soft Skills of Software Learning Development: the Psychological Dimensions of Computing and Security Behaviours}

\author{Matthew R Ivory}
%\orcidlink{0000-0002-5296-5897}}
\affiliation{
\institution{Supervised by Prof. J Towse, Prof. M Levine, Dr. M Sturdee, \& Prof. B Nuseibeh}
 \institution{Lancaster University}
 \city{Lancaster}
 \country{UK}}
 \email{matthew.ivory@lancaster.ac.uk}

\begin{abstract}
When writing software code, developers typically prioritise functionality over security, either consciously or unconsciously through biases and heuristics. This is often attributed to tangible pressures such as client requirements, but little is understood about the psychological dimensions affecting security behaviours. There is an increasing demand for understanding how psychological skills affect secure software development and to understand how these skills themselves are developed during the learning process.

This doctoral research explores this research space, with aims to identify important workplace-based skills for software developers; to identify and empirically investigate the soft skills behind these workplace skills in order to understand how soft skills can influence security behaviours; and, to identify ways to introduce and teach soft skills to computer science students to prepare the future generation of software developers.

The motivations behind this research are presented alongside the work plan. Three distinct phases are introduced, along with planned analyses. Phase one is currently in the data collection stage, with the second phase in planning. Prior relevant work is highlighted, and the paper concludes with a presentation of preliminary results and the planned next steps.
\end{abstract}

%%
%% The code below is generated by the tool at http://dl.acm.org/ccs.cfm.
%% Please copy and paste the code instead of the example below.
%%




\maketitle

\section{Introduction}

In this research, soft skills are defined as the psychological dimensions, or traits, that underpin behaviour \cite{capretzCallPromoteSoft2018}. Commonly, soft skills are synonymous with workplace-relevant transferable skills, including skills such as "teamwork" or "time management", and have been the focus of previous human factors research \cite{matturroSystematicMappingStudy2019, groeneveldSoftSkillsWhat2020, montandonWhatSkillsIT2021}. 
This current research seeks to go beyond these surface level traits, to identify the psychological dimensions that underpin transferable skills. 
In this body of research, transferable skills are referred to as "shallow skills", and soft skills are the underlying psychological dimensions. Shallow skills are referred to as such, because they provide little in the way of quantifiable skills, and their definitions often change depending on research context.
Shallow skills can be considered as the manifestation of soft skills, particularly in workplace situations. Soft skills are the more immutable, psychological aspects of behaviour.

Software development is the direct product of human interaction, created through the combination of cognitive abilities, social interactions and the unique culture of software development \cite{ahmedSoftSkillsSoftware2015, towseCaseUnderstandingSecure2020}. 
In recent years, the software industry has become aware of the significance of soft skills for successful software creation \cite{capretzCallPromoteSoft2018, matturroSystematicMappingStudy2019, montandonWhatSkillsIT2021}. 
By 2030, there is an anticipated 22\% increase in employment opportunities for software developers compared to an average 8\% increase across all other industries\footnote{\url{https://www.bls.gov/ooh/computer-and-information-technology/software-developers.htm}}, but a rising concern that graduates entering the workforce are lacking the necessary cognitive and social skills required for successful integration into the workplace \cite{liebenbergKnowledgeSkillsRequirements2014}. This issue has been evidenced in software security roles, with research indicating the most important skills required for security roles are not technical in nature, but are soft skills \cite{furnellAddressingCyberSecurity2020}. As a consequence, it is vital to identify the psychological traits required to successfully develop secure software.

Security in software is not a new concern, but the responsibility for security has changed over time. 
In 1999, Adams and Sasse \cite{adamsUsersAreNot1999} argued that software users were "not the enemy" and their fallible security behaviours were not their fault, but rather that of developers disregarding default user behaviour.

Similarly in 2008, Wurster and van Oorschot \cite{wursterDeveloperEnemy2008} posited that developers were "the enemy" and as they are the ones causing security issues, security should be removed from their responsibilities. They suggested the onus should be placed with API developers as they provide functionality (and security) to other developers. More recently, this sentiment about API developers was echoed by Green and Smith \cite{greenDevelopersAreNot2016}, who emphasised that developers typically focus on functionality and expect APIs to be secure by default. One issue with this argument is that it treats software developers as a homogeneous population with little security awareness, but expects API developers to be somehow more security conscious. API developers are as human as software developers, subsequently they are susceptible to the same cognitive and social biases \cite{oliveiraAPIBlindspotsWhy2018, brunBlindspotsPythonJava2021}. Rather than assigning responsibility to different groups, we should identify the psychological dimensions associated with good security behaviours and seek to promote these skills in software learning development.

\subsection{Motivation and Rationale}

The primary motivation is to understand how soft skills relate to software development, and how people's skills develop and exhibit in software development environments. Of particular interest are the behavioural changes exhibited by relative novices during skill development, compared to more experienced developers. What potentially incorrect, but intuitive actions are ultimately suppressed through experience? What habits are built and how do these originate? What soft skills are required for secure software development and how do these skills manifest and evolve?

In recent years, increased attention has turned towards the psychology of software developers, particularly in relation to security \cite{raufCaseAdaptiveSecurity2021}. Security vulnerabilities typically leverage psychological processes \cite{taylor-jacksonIncorporatingPsychologyCyber2020}, via cognitive processes (such as exploiting expected use cases), or through exploiting heuristic use. If adversaries exploit developers' behaviour, it is important to identify the soft skills involved and find ways in which these behaviours can be changed through psychological interventions, which can be taught to novice and experienced developers alike.

The project is motivated to provide practical impact through developing teaching materials for Computer Science courses. Incorporating psychological interventions into pedagogy will allow for the development of soft skills and security conscious behaviours in future generations of software developers.

\subsection{Contribution}

The main aim of this research is to understand how the software learning development process occurs and how behaviours change and evolve. By identifying these processes, we are better placed to encourage positive behavioural changes, resulting in more efficient code development. 
The project also seeks to investigate key soft skills that affect secure code production. As a result, psychological interventions can be developed for promoting better security practices. 
To practically encourage relevant behavioural changes in early-stage software developers (e.g. Computer Science students), pedagogical materials will be developed for education, with the aim for these to be incorporated into teaching practices.

\section{Research Questions}

RQ1: What soft skills are considered important for computing and security practices?

RQ2: How do soft skills evolve and develop in novice software developers with time and experience?

RQ3: How do software development behaviours evolve and change with experience?

RQ4: How can relevant soft skills be incorporated into pedagogical practice to promote security behaviours?

\section{Work Plan}

The research incorporates a breadth of analysis methods, including qualitative and quantitative approaches. This methodological pluralism allows for a broad range of information to be drawn from the data that would otherwise not be possible with a restricted methodology. The doctoral research will look at data collected through interviews, surveys, data scraping and behavioural studies. Analysis will be varied and include statistical modelling, natural language processing modelling, and qualitiative approaches, such as thematic analysis. Not only will it provide broader interpretation to findings within this space, it allows for stronger links to other work in similar research spaces.

\begin{figure}[htbp]
  \centering
 \includegraphics[width=0.47\textwidth]{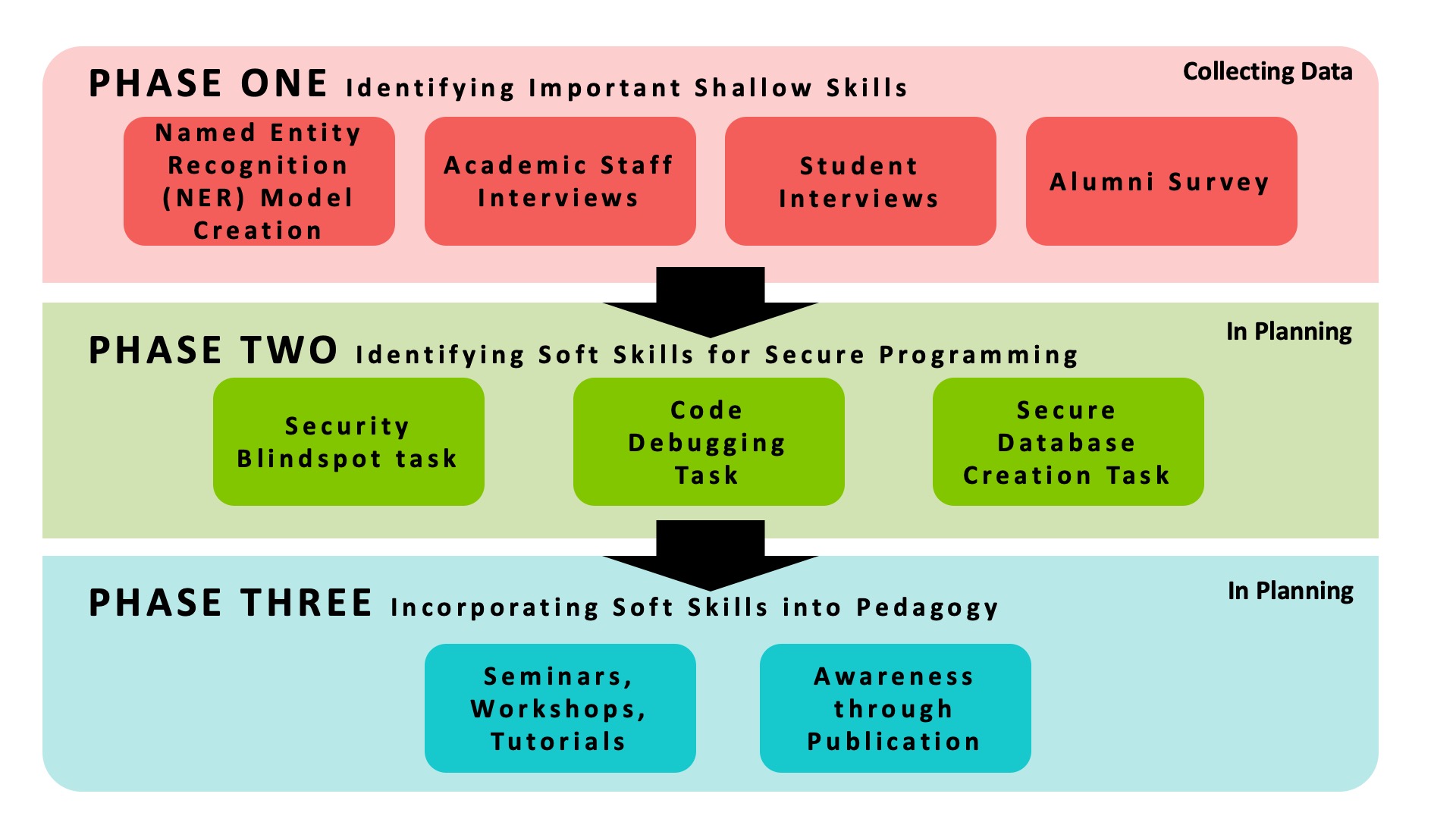}
 \caption{Outline of research phases within the planned work.}
   \label{fig:Figure 1}
\end{figure}

The research can be split into three phases, as illustrated by Figure \ref{fig:Figure 1}. The first phase identifies shallow skills as taught in Computer Science undergraduate courses. It seeks to understand how shallow skills are perceived by current students, staff and alumni.

The second phase will build directly on the first phase. By identifying soft skills linked to shallow skills through previous research, phase two aims to build relationships between secure coding behaviours and soft skills. This will be achieved through empirical, lab-based research involving manipulation and measuring of soft skills and programming tasks.

Finally, the third phase will focus on incorporating findings on soft skills into pedagogical materials. This phase will measure the effectiveness of teaching these ideas, with the aim to raise awareness and increase the understanding of the psychological traits of software developers.

A stand-alone study is also being conducted into cognitive reflection and risk perception in software developers and computer scientists, see section \ref{RiskPerception}. This will fit in with the phase two work.

Following open science practices, the research will include preregistrations, data sharing and reproducible analysis scripts. This will be managed through the use of the Open Science Framework\footnote{www.osf.io} and provision of Docker containers with reproducible workflows.

\subsection{Phase One: Identification of Shallow Skills}

The first phase is currently in the data collection stage. This phase is comprised of four research projects: an examination of core modules in computer science programmes as taken from university websites; academic staff interviews on how they view shallow skills being taught; longitudinal interviews with current students on how they develop their shallow skills over an academic year; and an alumni survey of computer science graduates and psychology graduates, collecting data on their perceived importance of shallow skills.

\subsubsection{Curriculum Examination.}

The online course information and core module descriptions for Computer Science and Psychology undergraduate courses were collected from eight UK universities belonging to the N8 research group (Durham, Lancaster, Leeds, Liverpool, Manchester, Newcastle, Sheffield and York).

To identify shallow skills in natural language texts, a named entity recognition (NER) model will be developed. Similar work has been created \cite{fareriSkillNERMiningMapping2021}, but without the granularity attempted here. In efforts to further understand covariance of shallow skills and language used around them, the NER model weights can be analysed further, including factor analysis to find highly correlated skills. A preregistration, providing details on the data collection can be found at \url{https://osf.io/qcw3n}.

\subsubsection{Alumni Survey.}

Lancaster University undergraduate alumni from Computer Science and Psychology were contacted to take part in a survey. Participants were asked to rank shallow skills for their importance in current employment. A Psychology sample were used as a comparative group, particularly when considering the less vocational nature of psychology undergraduate degrees (inferred from software development roles attained with a minimum education of a bachelor's degree\footnote{\url{https://nationalcareers.service.gov.uk/job-profiles/software-developer}}, compared to a minimum education of a postgraduate degree for most psychology roles\footnote{\url{https://nationalcareers.service.gov.uk/job-profiles/psychologist}}).

Data analysis will focus on loglinear models, correspondence analysis and exploratory factor analysis to identify the key shallow skills for computer science graduates compared to psychology graduates. The preregistration is found at \url{https://osf.io/5qb6a}.

\subsubsection{Interviews with Staff and Students.}

These two projects are planned, and interview schedules will be arranged for times when teaching volume is low for staff, and longitudinal student surveys will begin in line with the start of an academic year.

Staff interviews will look to identify soft skills considered important by teaching staff and how these are conveyed to students in teaching materials.
Student interviews will be conducted over the course of the academic year, following the same students to identify the way in which they recognise and develop shallow skills. Analysis for both interview studies will use thematic and content analysis to extract relevant information.

\subsubsection{Data Analysis of Phase One.}

\begin{figure}[htbp]
  \centering
 \includegraphics[width=0.47\textwidth]{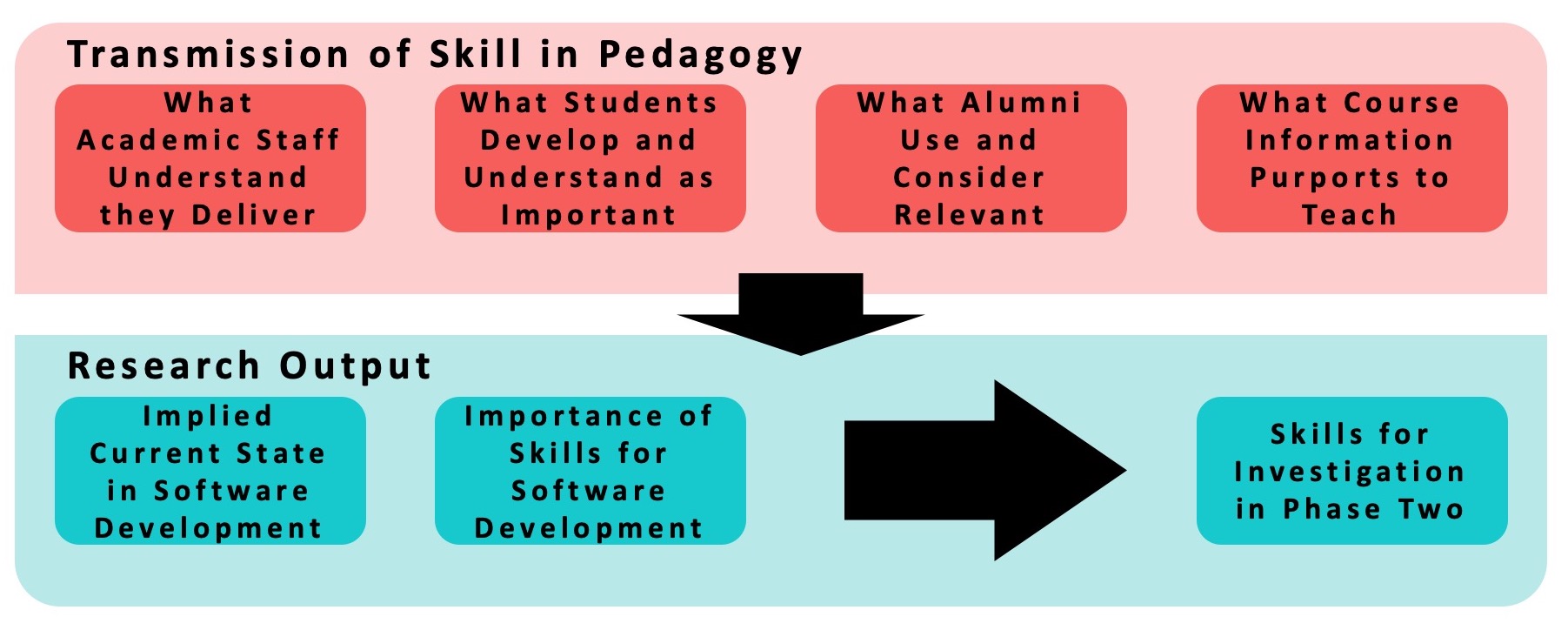}
 \caption{Transmission of shallow skills can be expected to develop whereby teaching staffs' understanding of which skills are important feed into the skills students pick up on, which are reflected in alumni use of these skills.}
   \label{fig:Figure 2}
\end{figure}

The results from the individual projects in phase one can be cross-examined to identify areas of shallow skills that are of most interest. 
The combined data can be used to understand the transmission of ideas from academic staff to students to what they take into the workplace, (see figure \ref{fig:Figure 2}). Understanding the development of these skills and their importance can be used in the second phase. Analysis of data is in planning.

\emph{Goal: to identify the shallow skills considered as important within the transmission of skills in the pedagogical process. Measured through various quantitative and qualitative methods.}

\subsection{Phase Two: Behavioural Studies}

The second phase is in early design stages, but will focus on empirical behavioural research, based on the findings from phase one. The exact soft skills to be included is dependent on phase one findings, as it is important to focus on the skills that are most likely to have the biggest effect on coding behaviours.

One study will investigate API blindspots, which can be defined as a misunderstanding or misrepresentation of API function security, resulting in vulnerabilities \cite{oliveiraAPIBlindspotsWhy2018}. Using Python snippets from Brun et al (2021) \cite{brunBlindspotsPythonJava2021}, and measuring soft skills through cognitive tasks (e.g. the cognitive reflection test \cite{frederickCognitiveReflectionDecision2005}), relationships can be drawn between soft skills and API blindspot awareness.

Similar studies, using different programming paradigms (such as code debugging, or secure password database creation) will also be used. It is important to understand the stability of soft skills across a range of development and security-related tasks.

\subsubsection{Data Analysis of Phase Two.}

Data analysis for phase two experiments will be predominantly quantitative, using mixed effects models for group comparisons. These can be used to measure relationships between soft skills and security behaviours. Preregistrations will be published in due course.

\emph{Goal: to identify and measure the effect of soft skills on security behaviours. Measured through mixed effect modelling and group comparisons.}

\subsection{Phase Three: Inclusion in Pedagogy}

For the final phase, the focus will be on the development of pedagogical materials for introducing students to the soft skills necessary for secure programming.
This phase has not yet reached planning, as it relies on the work of phase two to be near completion.
This will be achieved through seminars or workshops as methods to introduce the soft skills, to encourage students to engage with the psychology behind software development.
Effectiveness of sessions will likely be measured through participant feedback.

To further disseminate research findings and promote inclusion of soft skills into current pedagogical materials, engagement through publication will be pursued. By raising awareness of research through publication, conferences and posters, along with the provision of basic materials for others to work with, phase three looks to create a meaningful impact in the domain of software learning development. 

\emph{Goal: to develop and deliver teaching materials in order to promote soft skills within computer science curricula. Measured through student engagement and feedback.}

\subsection{Risk Perception and Cognitive Reflection} \label{RiskPerception}

In this individual differences study aligned with phase two research, groups of professional software developers and computer science students were compared regarding risk perception in software. Participants completed a cognitive reflection test, a risk-oriented decision task, and answered qualitative questions about how they understand risk in software development.

Cognitive reflection is a person's ability to inhibit intuitive responses in favour of more reflective responses, indicating their skill in reflective thinking in search of a correct answer.
Cognitive reflection was measured through the Cognitive Reflection Test (CRT) \cite{frederickCognitiveReflectionDecision2005}. 
This is a three question test, including items such as, "A paperclip and an elastic band cost £1.10 in total. The elastic band costs £1 more than the paperclip. How much does the paperclip cost?" The intuitive answer is 10 pence, but upon reflection the correct answer is 5 pence. The risk-orientation task focussed on how participants view susceptibility of themselves and the "average developer" when considering security vulnerabilities as listed by OWASP (e.g. SQL injection). Data analysis is in progress. The preregistration can be found at: \url{https://osf.io/zbqe4}. 

\subsubsection{Data Analysis of Risk Perception Study.}

Data will be analysed through quantitative measures, such as linear modelling, along with more qualitative methods, including thematic and content analyses.

\emph{Goal: to identify potential relationships between risk-related behaviours in software development and cognitive reflection. Measured through linear modelling and group comparisons.}

\subsection{Validity Threats and Controls}

Validity threats to the research are broadly discussed, relevant to the project overall. More granular considerations are included in preregistration documents.

One key threat is the consideration of software developers as a population. It is easy to treat developers as a homogeneous population who demonstrate similar characteristics, subsequently making approaches to promoting security behaviours intolerant to variance within the population. This can be controlled through mixed effect models, where population characteristics can be included in the analysis to identify the effect these have on behaviours. 

This is a relatively new research field \cite{raufCaseAdaptiveSecurity2021}, so much of the planned research is exploratory. This can often lead to a series of analysis methods being used, increasing type I errors. To control for this, preregistration procedures are published prior to data collection. Open data and reproducible analysis scripts will be uploaded following study completion, to allow replication and to confirm findings.

Using a range of methodologies, as highlighted in the work plan, may result in a trade-off between breadth of analysis and depth of analysis. To control for this, analysis plans and research choices will be well considered through the use of preregistration documents. By considering methodologies prior to execution, the connection between the studies within the wider research can be well justified. 

Another threat to validity is the generalisation to different programming languages or work cultures. Not all languages have similar structures, and differences have been shown in security awareness between Java and Python APIs \cite{brunBlindspotsPythonJava2021}. This can be controlled for by acknowledging that results may only apply to a single language. By focussing on Python, which is the most popular language\footnote{\url{https://www.tiobe.com/tiobe-index/python/}}, findings will have relevance to many developers. The inclusion of preregistration documents, materials and analyses will also allow for replications, either directly or conceptually.

The tasks used in the second phase are designed to provide consistency across participants, reducing task variance and improving statistical power. This comes at a cost, which is that the tasks are less industry-specific, reducing the validity. This PhD research is specifically focussed on the software learning process, and work beyond the PhD may look into more industry specific tasks, or applying similar research to different programming languages.

\section{Relevant Prior Work}

In this section the current literature relevant to the research is discussed. This is not an exhaustive literature review, but aims to identify key research influencing the doctoral research.

\subsection{Phase One}

In phase one, key research identified shallow skills in software development, such as Matturro et al. (2019) \cite{matturroSystematicMappingStudy2019}, who conducted a systematic analysis and identified 23 separate skills. Similarly, Stevens and Norman (2016) looked at job adverts to identify the most important shallow skills for developers \cite{stevensIndustryExpectationsSoft2016}. These research papers provided context for the important shallow skills.

Groeneveld et al. \cite{groeneveldSoftSkillsWhat2020} analysed computer science curricula for modules that taught shallow skills explicitly, but did not look into the implicitly taught skills in all modules. This motivated the investigation of the course curricula for text relevant to shallow skills.

Finding an absence of research that provided associations between shallow skills and soft skills is the motivation for phase one. The literature search for phase one has found little evidence of work associating security awareness and soft skills.

\subsection{Phase Two}

In the second phase, a series of work has been carried out concerning API blindspots and developers' use of heuristics when evaluating software code. Oliveira et al. (2018) \cite{oliveiraAPIBlindspotsWhy2018} highlighted this issue with Java puzzles, finding that security blindspots in code snippets were difficult to identify, possibly due to developers' expectation of APIs being secure as default.
Brun et al. (2021) \cite{brunBlindspotsPythonJava2021} followed this work with a replication using Python code. They found that developers who exhibited better long term memory recall were more successful in solving puzzles with blindspots. They found that short term memory, memory span and episodic memory had no effect on solving the puzzles. Other works that touch on psychology in security include Hallett et al (2021) \cite{hallettThisThatNothing2021}, where boosting security awareness through requiring planning promoted a small effect on security, and Shreeve et al (2020) \cite{shreeveIfMrBlue2020} who identified decision making processes related to cybersecurity.

\subsection{Phase Three}

For the third phase, Taylor-Jackson et al. (2020) \cite{taylor-jacksonIncorporatingPsychologyCyber2020} advocated including psychology in security education, particularly when considering that vulnerabilities are often psychological in nature (e.g. phishing, API blindspots). They discuss the benefits of exposing computer scientists to the different ideas and styles of thinking found within psychology. 
There are also wider calls for inclusion of soft skills in university education \cite{guerra-baezPanoramicReviewSoft2019}. It is important that the findings from the first two research phases are used for positive impact and one immediate way to achieve this, is to answer the calls for increasing soft skill teachings in cybersecurity courses to benefit future software developers.

\subsection{Risk Perception}

For this study, key items are papers on cognitive reflection by Frederick (2005) \cite{frederickCognitiveReflectionDecision2005} and Thomson and Oppenheimer (2016) \cite{thomsonInvestigatingAlternateForm2016}. Combined with the understanding that developers are often not the most security conscious, as highlighted by Acar et al (2017) \cite{acarDevelopersNeedSupport2017}, it is clear that the understanding of risk by developers in a software context is poorly understood in relation to cognitive measures.

\section{Current Status}
\subsection{Early Results Analysis}

Some of the preliminary results from the risk perception study (section \ref{RiskPerception}) are mentioned here. The third hypothesis stated in the preregistration is examined here, "Mean scores closer to zero on the novel OWASP risk task will be found with higher scores of cognitive reflection". Data from 143 (70 students, 73 developers) participants is used.

The OWASP task is a measure devised for this study where participants were asked to respond to two sets of questions, the first asking about the percentage of web applications they believe to be created by others that suffer from one of the top five OWASP vulnerabilities (injection flaws, broken authentication, sensitive data exposure, XML External Entity and broken access control). Then following a separation task, participants were then asked to rate the percentage of web applications that they had developed that suffered from the same vulnerabilities. Scores closer to 100 indicate high optimism that they do not produce flawed products, scores near 0 indicate similar levels of flaws in both their own and other people's products, and scores approaching -100 indicate beliefs that their own work is highly susceptible to these vulnerabilities.

To test the hypothesis above, a linear regression was run to see whether CRT scores significantly predicted OWASP vulnerability scores and whether this differed between the two populations.
The model formula was "Vulnerability $\sim$ CRT". The overall regression was statistically significant (R$^2$ = .05, F(3, 139) = 15.13, \emph{p} = .017. Estimates, values and significance of model items can be seen in Table \ref{tab:Table 1}. Despite significant terms in the model summary, the variance explained by the model is negligible ($\sim$5\%) and further models will need to be developed to explain more variance in these scores.

\begin{figure}[htbp]
  \centering
 \includegraphics[width=0.4\textwidth]{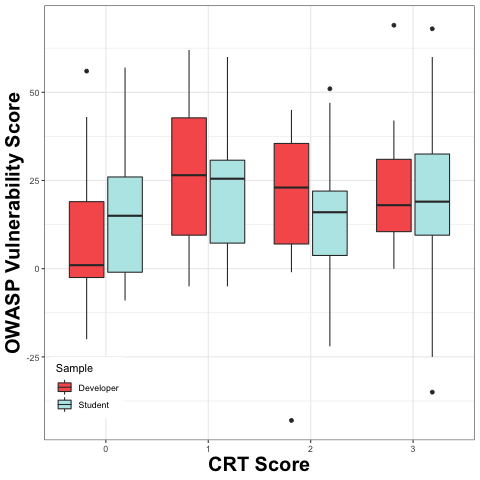}
 \caption{Box plot of mean OWASP vulnerability scores by CRT score split by population.}
   \label{fig:Figure 4}
\end{figure}

Figure \ref{fig:Figure 4} shows the distribution of vulnerability scores for each level of CRT score as a box plot. Post-hoc Tukey tests identified no significant differences between any of the groups with all \emph{p} > .05, except for those who scored zero and those who correctly scored one, adjusted \emph{p} = .014. This indicates that there is little significant relationship between CRT scores and results on the novel OWASP risk task.

\begin{table}
  \caption{Coefficients, t-values and \emph{p}-values for the linear regression of CRT predicting OWASP vulnerability}
  \label{tab:Table 1}
  \begin{tabular}{lrrrl}
\toprule
  & Estimate & Std. Error & t value & \emph{p}*\\
\midrule
Intercept & 11.00  & 3.13 &  3.52 &    <.001*** \\
CRTscore1 & 14.08  & 4.59 &  3.07 & .003**  \\
CRTscore2 & 5.61   & 4.86 &  1.15 & .250\\
CRTscore3 & 10.67  & 4.91 &  2.17 & .031*\\
\bottomrule
  \end{tabular}
  {\bigskip \raggedright *Significant alpha values of <.001 indicated by ***\par}
  \vspace{-5mm}
\end{table}

What is noted with the Vulnerability scores, and can be seen in Figure \ref{fig:Figure 4}, is that most scores on the OWASP task, regardless of CRT scores, are around or above 0. One-way t-tests on the Vulnerability scores were run for both the developer and the student samples. In the Developer sample (mean score = 17.48), the scores were significantly higher than 0, t(69) = 7.16, \emph{p} < .001. Similarly for the student sample (mean score = 18.94), scores were significantly higher than 0, t(69) = 7.26, \emph{p} < .001.

This finding is indicative of optimism bias \cite{sharotOptimismBias2011}, suggesting that both professionals and student developers consider themselves to be better than average at preventing these OWASP-listed security issues. A score of zero would indicate respondents understand they were average, but higher scores suggest an over-optimistic outlook on their own abilities, which could lead to a more relaxed view on these security issues. These findings will be further developed and discussed in future publications.

\subsection{Next Steps}

In the short term, the next steps are to continue with phase one data collection, and planning of phase two. It is intended that the research will progress according to the research phases outlined above. Following the completion of the doctoral work, future work would include the investigation of psychological interventions for developing soft skills and measuring their impact on security behaviours in longitudinal research.

The steps beyond the PhD research as outlined above is to focus on the skills that explain the largest variance in secure coding behaviours, and seek to identify the best ways to promote continued, stable use of these behaviours as opposed to short-term changes (such as those achieved through nudging, e.g. IDE pop-ups serving as reminders to look for blindspots).

\section{Conclusion}

This paper provides an overview of the planned work within the PhD research titled "The Soft Skills of Software Learning Development: the Psychological Dimensions of Computing and Security Behaviours". The research is diverse in both aims and processes, ranging from thematic analysis of interviews, to modelling relationships between psychological dimensions and security issues, to incorporating the findings into pedagogy. 

This project seeks to investigate the software learning development process; to better understand the behavioural changes and soft skill development of both computing and security behaviours. By identifying these changes, and when and how they develop, we can seek to promote these changes earlier in the learning cycle, allowing for more effective learning and encouraging positive behaviours for software development. In doing so, not only can we exhibit greater awareness of the psychology behind secure software development, we can develop interventions for encouraging these secure behaviours, reducing the likelihood of these security vulnerabilities. All public preregistrations, published data, analyses, and links to further research outputs will be accessible from \url{https://osf.io//v93zt}.

\bibliographystyle{ACM-Reference-Format}
\bibliography{ease2022-65-bib}

%%% -*-BibTeX-*-
%%% Do NOT edit. File created by BibTeX with style
%%% ACM-Reference-Format-Journals [18-Jan-2012].

\begin{thebibliography}{24}

%%% ====================================================================
%%% NOTE TO THE USER: you can override these defaults by providing
%%% customized versions of any of these macros before the \bibliography
%%% command.  Each of them MUST provide its own final punctuation,
%%% except for \shownote{}, \showDOI{}, and \showURL{}.  The latter two
%%% do not use final punctuation, in order to avoid confusing it with
%%% the Web address.
%%%
%%% To suppress output of a particular field, define its macro to expand
%%% to an empty string, or better, \unskip, like this:
%%%
%%% \newcommand{\showDOI}[1]{\unskip}   % LaTeX syntax
%%%
%%% \def \showDOI #1{\unskip}           % plain TeX syntax
%%%
%%% ====================================================================

\ifx \showCODEN    \undefined \def \showCODEN     #1{\unskip}     \fi
\ifx \showDOI      \undefined \def \showDOI       #1{#1}\fi
\ifx \showISBNx    \undefined \def \showISBNx     #1{\unskip}     \fi
\ifx \showISBNxiii \undefined \def \showISBNxiii  #1{\unskip}     \fi
\ifx \showISSN     \undefined \def \showISSN      #1{\unskip}     \fi
\ifx \showLCCN     \undefined \def \showLCCN      #1{\unskip}     \fi
\ifx \shownote     \undefined \def \shownote      #1{#1}          \fi
\ifx \showarticletitle \undefined \def \showarticletitle #1{#1}   \fi
\ifx \showURL      \undefined \def \showURL       {\relax}        \fi
% The following commands are used for tagged output and should be
% invisible to TeX
\providecommand\bibfield[2]{#2}
\providecommand\bibinfo[2]{#2}
\providecommand\natexlab[1]{#1}
\providecommand\showeprint[2][]{arXiv:#2}

\bibitem[\protect\citeauthoryear{Acar, Stransky, Wermke, Weir, Mazurek, and
  Fahl}{Acar et~al\mbox{.}}{2017}]%
        {acarDevelopersNeedSupport2017}
\bibfield{author}{\bibinfo{person}{Yasemin Acar}, \bibinfo{person}{Christian
  Stransky}, \bibinfo{person}{Dominik Wermke}, \bibinfo{person}{Charles Weir},
  \bibinfo{person}{Michelle~L. Mazurek}, {and} \bibinfo{person}{Sascha Fahl}.}
  \bibinfo{year}{2017}\natexlab{}.
\newblock \showarticletitle{Developers {{Need Support}}, {{Too}}: {{A Survey}}
  of {{Security Advice}} for {{Software Developers}}}. In
  \bibinfo{booktitle}{\emph{2017 {{IEEE Cybersecurity Development}}
  ({{SecDev}})}}. \bibinfo{publisher}{{IEEE}}, \bibinfo{address}{{Cambridge,
  MA, USA}}, \bibinfo{pages}{22--26}.
\newblock
\urldef\tempurl%
\url{https://doi.org/10.1109/SecDev.2017.17}
\showDOI{\tempurl}


\bibitem[\protect\citeauthoryear{Adams and Sasse}{Adams and Sasse}{1999}]%
        {adamsUsersAreNot1999}
\bibfield{author}{\bibinfo{person}{Anne Adams} {and}
  \bibinfo{person}{Martina~Angela Sasse}.} \bibinfo{year}{1999}\natexlab{}.
\newblock \showarticletitle{Users Are Not the Enemy}.
\newblock \bibinfo{journal}{\emph{Commun. ACM}} \bibinfo{volume}{42},
  \bibinfo{number}{12} (\bibinfo{date}{Dec.} \bibinfo{year}{1999}),
  \bibinfo{pages}{40--46}.
\newblock
\showISSN{0001-0782, 1557-7317}
\urldef\tempurl%
\url{https://doi.org/10.1145/322796.322806}
\showDOI{\tempurl}


\bibitem[\protect\citeauthoryear{Ahmed, Capretz, Bouktif, and Campbell}{Ahmed
  et~al\mbox{.}}{2015}]%
        {ahmedSoftSkillsSoftware2015}
\bibfield{author}{\bibinfo{person}{Faheem Ahmed},
  \bibinfo{person}{Luiz~Fernando Capretz}, \bibinfo{person}{Salah Bouktif},
  {and} \bibinfo{person}{Piers Campbell}.} \bibinfo{year}{2015}\natexlab{}.
\newblock \showarticletitle{Soft {{Skills}} and {{Software Development}}: {{A
  Reflection}} from the {{Software Industry}}}.
\newblock \bibinfo{journal}{\emph{International Journal of Information
  Processing and Management}} \bibinfo{volume}{4}, \bibinfo{number}{3}
  (\bibinfo{date}{July} \bibinfo{year}{2015}), \bibinfo{pages}{171--191}.
\newblock
\urldef\tempurl%
\url{https://doi.org/10.4156/ijipm.vol14.issue3.17}
\showDOI{\tempurl}
\showeprint[arxiv]{1507.06873}


\bibitem[\protect\citeauthoryear{Brun, Lin, Somerville, Myers, and Ebner}{Brun
  et~al\mbox{.}}{2021}]%
        {brunBlindspotsPythonJava2021}
\bibfield{author}{\bibinfo{person}{Yuriy Brun}, \bibinfo{person}{Tian Lin},
  \bibinfo{person}{Jessie~Elise Somerville}, \bibinfo{person}{Elisha Myers},
  {and} \bibinfo{person}{Natalie~C. Ebner}.} \bibinfo{year}{2021}\natexlab{}.
\newblock \showarticletitle{Blindspots in {{Python}} and {{Java APIs Result}}
  in {{Vulnerable Code}}}.
\newblock \bibinfo{journal}{\emph{arXiv:2103.06091 [cs]}}
  (\bibinfo{date}{March} \bibinfo{year}{2021}).
\newblock
\showeprint[arxiv]{2103.06091}~[cs]


\bibitem[\protect\citeauthoryear{Capretz and Ahmed}{Capretz and Ahmed}{2018}]%
        {capretzCallPromoteSoft2018}
\bibfield{author}{\bibinfo{person}{Luiz~Fernando Capretz} {and}
  \bibinfo{person}{Faheem Ahmed}.} \bibinfo{year}{2018}\natexlab{}.
\newblock \showarticletitle{A {{Call}} to {{Promote Soft Skills}} in {{Software
  Engineering}}}.
\newblock \bibinfo{journal}{\emph{Psychology and Cognitive Sciences - Open
  Journal}} \bibinfo{volume}{4}, \bibinfo{number}{1} (\bibinfo{date}{Aug.}
  \bibinfo{year}{2018}), \bibinfo{pages}{e1--e3}.
\newblock
\showISSN{2380727X}
\urldef\tempurl%
\url{https://doi.org/10.17140/PCSOJ-4-e011}
\showDOI{\tempurl}
\showeprint[arxiv]{1901.01819}


\bibitem[\protect\citeauthoryear{Fareri, Melluso, Chiarello, and
  Fantoni}{Fareri et~al\mbox{.}}{2021}]%
        {fareriSkillNERMiningMapping2021}
\bibfield{author}{\bibinfo{person}{Silvia Fareri}, \bibinfo{person}{Nicola
  Melluso}, \bibinfo{person}{Filippo Chiarello}, {and}
  \bibinfo{person}{Gualtiero Fantoni}.} \bibinfo{year}{2021}\natexlab{}.
\newblock \showarticletitle{{{SkillNER}}: {{Mining}} and {{Mapping Soft
  Skills}} from Any {{Text}}}.
\newblock \bibinfo{journal}{\emph{Expert Systems with Applications}}
  \bibinfo{volume}{184} (\bibinfo{date}{Dec.} \bibinfo{year}{2021}),
  \bibinfo{pages}{115544}.
\newblock
\showISSN{09574174}
\urldef\tempurl%
\url{https://doi.org/10.1016/j.eswa.2021.115544}
\showDOI{\tempurl}
\showeprint[arxiv]{2101.11431}


\bibitem[\protect\citeauthoryear{Frederick}{Frederick}{2005}]%
        {frederickCognitiveReflectionDecision2005}
\bibfield{author}{\bibinfo{person}{Shane Frederick}.}
  \bibinfo{year}{2005}\natexlab{}.
\newblock \showarticletitle{Cognitive Reflection and Decision Making}.
\newblock \bibinfo{journal}{\emph{Journal of Economic perspectives}}
  \bibinfo{volume}{19}, \bibinfo{number}{4} (\bibinfo{year}{2005}),
  \bibinfo{pages}{25--42}.
\newblock
\showISSN{0895-3309}


\bibitem[\protect\citeauthoryear{Furnell and Bishop}{Furnell and
  Bishop}{2020}]%
        {furnellAddressingCyberSecurity2020}
\bibfield{author}{\bibinfo{person}{Steven Furnell} {and} \bibinfo{person}{Matt
  Bishop}.} \bibinfo{year}{2020}\natexlab{}.
\newblock \showarticletitle{Addressing Cyber Security Skills: The Spectrum, Not
  the Silo}.
\newblock \bibinfo{journal}{\emph{Computer Fraud \& Security}}
  \bibinfo{volume}{2020}, \bibinfo{number}{2} (\bibinfo{date}{Feb.}
  \bibinfo{year}{2020}), \bibinfo{pages}{6--11}.
\newblock
\showISSN{1361-3723}
\urldef\tempurl%
\url{https://doi.org/10.1016/S1361-3723(20)30017-8}
\showDOI{\tempurl}


\bibitem[\protect\citeauthoryear{Green and Smith}{Green and Smith}{2016}]%
        {greenDevelopersAreNot2016}
\bibfield{author}{\bibinfo{person}{Matthew Green} {and}
  \bibinfo{person}{Matthew Smith}.} \bibinfo{year}{2016}\natexlab{}.
\newblock \showarticletitle{Developers Are {{Not}} the {{Enemy}}!: {{The Need}}
  for {{Usable Security APIs}}}.
\newblock \bibinfo{journal}{\emph{IEEE Security Privacy}} \bibinfo{volume}{14},
  \bibinfo{number}{5} (\bibinfo{date}{Sept.} \bibinfo{year}{2016}),
  \bibinfo{pages}{40--46}.
\newblock
\showISSN{1558-4046}
\urldef\tempurl%
\url{https://doi.org/10.1109/MSP.2016.111}
\showDOI{\tempurl}


\bibitem[\protect\citeauthoryear{Groeneveld, Becker, and Vennekens}{Groeneveld
  et~al\mbox{.}}{2020}]%
        {groeneveldSoftSkillsWhat2020}
\bibfield{author}{\bibinfo{person}{Wouter Groeneveld},
  \bibinfo{person}{Brett~A. Becker}, {and} \bibinfo{person}{Joost Vennekens}.}
  \bibinfo{year}{2020}\natexlab{}.
\newblock \showarticletitle{Soft {{Skills}}: {{What}} Do {{Computing Program
  Syllabi Reveal About Non-Technical Expectations}} of {{Undergraduate
  Students}}?}. In \bibinfo{booktitle}{\emph{Proceedings of the 2020 {{ACM
  Conference}} on {{Innovation}} and {{Technology}} in {{Computer Science
  Education}}}} \emph{(\bibinfo{series}{{{ITiCSE}} '20})}.
  \bibinfo{publisher}{{Association for Computing Machinery}},
  \bibinfo{address}{{New York, NY, USA}}, \bibinfo{pages}{287--293}.
\newblock
\showISBNx{978-1-4503-6874-2}
\urldef\tempurl%
\url{https://doi.org/10.1145/3341525.3387396}
\showDOI{\tempurl}


\bibitem[\protect\citeauthoryear{{Guerra-B{\'a}ez}}{{Guerra-B{\'a}ez}}{2019}]%
        {guerra-baezPanoramicReviewSoft2019}
\bibfield{author}{\bibinfo{person}{Sandra~Patricia {Guerra-B{\'a}ez}}.}
  \bibinfo{year}{2019}\natexlab{}.
\newblock \showarticletitle{A Panoramic Review of Soft Skills Training in
  University Students}.
\newblock \bibinfo{journal}{\emph{Psicologia Escolar e Educacional}}
  \bibinfo{volume}{23} (\bibinfo{year}{2019}), \bibinfo{pages}{1--10}.
\newblock
\showISSN{2175-3539}
\urldef\tempurl%
\url{https://doi.org/10.1590/2175-35392019016464}
\showDOI{\tempurl}


\bibitem[\protect\citeauthoryear{Hallett, Patnaik, Shreeve, and Rashid}{Hallett
  et~al\mbox{.}}{2021}]%
        {hallettThisThatNothing2021}
\bibfield{author}{\bibinfo{person}{Joseph Hallett}, \bibinfo{person}{Nikhil
  Patnaik}, \bibinfo{person}{Benjamin Shreeve}, {and} \bibinfo{person}{Awais
  Rashid}.} \bibinfo{year}{2021}\natexlab{}.
\newblock \showarticletitle{``{{Do}} This! {{Do}} That!, {{And}} Nothing Will
  Happen'' {{Do}} Specifications Lead to Securely Stored Passwords?}. In
  \bibinfo{booktitle}{\emph{Proceedings of the 43rd {{International
  Conference}} on {{Software Engineering}} ({{ICSE}} '21)}}.
  \bibinfo{publisher}{{IEEE}}, \bibinfo{address}{{Madrid, Spain}},
  \bibinfo{pages}{486--498}.
\newblock
\urldef\tempurl%
\url{https://doi.org/10.1109/ICSE43902.2021.00053}
\showDOI{\tempurl}


\bibitem[\protect\citeauthoryear{Liebenberg, Huisman, and Mentz}{Liebenberg
  et~al\mbox{.}}{2014}]%
        {liebenbergKnowledgeSkillsRequirements2014}
\bibfield{author}{\bibinfo{person}{Janet Liebenberg}, \bibinfo{person}{Magda
  Huisman}, {and} \bibinfo{person}{Elsa Mentz}.}
  \bibinfo{year}{2014}\natexlab{}.
\newblock \showarticletitle{Knowledge and {{Skills Requirements}} for
  {{Software Developer Students}}}.
\newblock \bibinfo{journal}{\emph{International Journal of Social, Behavioral,
  Educational, Economic, Business and Industrial Engineering}}
  \bibinfo{volume}{8}, \bibinfo{number}{8} (\bibinfo{year}{2014}),
  \bibinfo{pages}{6}.
\newblock


\bibitem[\protect\citeauthoryear{Matturro, Raschetti, and Font{\'a}n}{Matturro
  et~al\mbox{.}}{2019}]%
        {matturroSystematicMappingStudy2019}
\bibfield{author}{\bibinfo{person}{Gerardo Matturro},
  \bibinfo{person}{Florencia Raschetti}, {and} \bibinfo{person}{Carina
  Font{\'a}n}.} \bibinfo{year}{2019}\natexlab{}.
\newblock \showarticletitle{A {{Systematic Mapping Study}} on {{Soft Skills}}
  in {{Software Engineering}}}.
\newblock \bibinfo{journal}{\emph{Journal of Universal Computer Science}}
  \bibinfo{volume}{25}, \bibinfo{number}{1} (\bibinfo{year}{2019}),
  \bibinfo{pages}{26}.
\newblock


\bibitem[\protect\citeauthoryear{Montandon, Politowski, Silva, Valente,
  Petrillo, and Gu{\'e}h{\'e}neuc}{Montandon et~al\mbox{.}}{2021}]%
        {montandonWhatSkillsIT2021}
\bibfield{author}{\bibinfo{person}{Jo{\~a}o~Eduardo Montandon},
  \bibinfo{person}{Cristiano Politowski}, \bibinfo{person}{Luciana~Lourdes
  Silva}, \bibinfo{person}{Marco~Tulio Valente}, \bibinfo{person}{Fabio
  Petrillo}, {and} \bibinfo{person}{Yann-Ga{\"e}l Gu{\'e}h{\'e}neuc}.}
  \bibinfo{year}{2021}\natexlab{}.
\newblock \showarticletitle{What Skills Do {{IT}} Companies Look for in New
  Developers? {{A}} Study with {{Stack Overflow}} Jobs}.
\newblock \bibinfo{journal}{\emph{Information and Software Technology}}
  \bibinfo{volume}{129} (\bibinfo{date}{Jan.} \bibinfo{year}{2021}),
  \bibinfo{pages}{106429}.
\newblock
\showISSN{0950-5849}
\urldef\tempurl%
\url{https://doi.org/10.1016/j.infsof.2020.106429}
\showDOI{\tempurl}


\bibitem[\protect\citeauthoryear{Oliveira, Lin, Rahman, Akefirad, Ellis, Perez,
  Bobhate, DeLong, Cappos, and Brun}{Oliveira et~al\mbox{.}}{2018}]%
        {oliveiraAPIBlindspotsWhy2018}
\bibfield{author}{\bibinfo{person}{Daniela~Seabra Oliveira},
  \bibinfo{person}{Tian Lin}, \bibinfo{person}{Muhammad~Sajidur Rahman},
  \bibinfo{person}{Rad Akefirad}, \bibinfo{person}{Donovan Ellis},
  \bibinfo{person}{Eliany Perez}, \bibinfo{person}{Rahul Bobhate},
  \bibinfo{person}{Lois~A DeLong}, \bibinfo{person}{Justin Cappos}, {and}
  \bibinfo{person}{Yuriy Brun}.} \bibinfo{year}{2018}\natexlab{}.
\newblock \showarticletitle{\{\vphantom\}{{API}}\vphantom\{\} {{Blindspots}}:
  {{Why Experienced Developers Write Vulnerable Code}}}. In
  \bibinfo{booktitle}{\emph{Fourteenth {{Symposium}} on {{Usable Privacy}} and
  {{Security}} (\{\vphantom\}{{SOUPS}}\vphantom\{\} 2018)}}.
  \bibinfo{publisher}{{USENIX Association}}, \bibinfo{address}{{Baltimore, MD,
  USA}}, \bibinfo{pages}{315--328}.
\newblock
\showISBNx{1-939133-10-6}


\bibitem[\protect\citeauthoryear{Rauf, Petre, Tun, Lopez, Lunn, Van Der~Linden,
  Towse, Sharp, Levine, Rashid, and Nuseibeh}{Rauf et~al\mbox{.}}{2021}]%
        {raufCaseAdaptiveSecurity2021}
\bibfield{author}{\bibinfo{person}{Irum Rauf}, \bibinfo{person}{Marian Petre},
  \bibinfo{person}{Thein Tun}, \bibinfo{person}{Tamara Lopez},
  \bibinfo{person}{Paul Lunn}, \bibinfo{person}{Dirk Van Der~Linden},
  \bibinfo{person}{John Towse}, \bibinfo{person}{Helen Sharp},
  \bibinfo{person}{Mark Levine}, \bibinfo{person}{Awais Rashid}, {and}
  \bibinfo{person}{Bashar Nuseibeh}.} \bibinfo{year}{2021}\natexlab{}.
\newblock \showarticletitle{The {{Case}} for {{Adaptive Security
  Interventions}}}.
\newblock \bibinfo{journal}{\emph{ACM Transactions on Software Engineering and
  Methodology}} \bibinfo{volume}{31}, \bibinfo{number}{1}
  (\bibinfo{date}{Sept.} \bibinfo{year}{2021}), \bibinfo{pages}{9:1--9:52}.
\newblock
\showISSN{1049-331X}
\urldef\tempurl%
\url{https://doi.org/10.1145/3471930}
\showDOI{\tempurl}


\bibitem[\protect\citeauthoryear{Sharot}{Sharot}{2011}]%
        {sharotOptimismBias2011}
\bibfield{author}{\bibinfo{person}{Tali Sharot}.}
  \bibinfo{year}{2011}\natexlab{}.
\newblock \showarticletitle{The Optimism Bias}.
\newblock \bibinfo{journal}{\emph{Current Biology}} \bibinfo{volume}{21},
  \bibinfo{number}{23} (\bibinfo{date}{Dec.} \bibinfo{year}{2011}),
  \bibinfo{pages}{R941--R945}.
\newblock
\showISSN{0960-9822}
\urldef\tempurl%
\url{https://doi.org/10.1016/j.cub.2011.10.030}
\showDOI{\tempurl}


\bibitem[\protect\citeauthoryear{Shreeve, Hallett, Edwards, Anthonysamy, Frey,
  and Rashid}{Shreeve et~al\mbox{.}}{2020}]%
        {shreeveIfMrBlue2020}
\bibfield{author}{\bibinfo{person}{Benjamin Shreeve}, \bibinfo{person}{Joseph
  Hallett}, \bibinfo{person}{Matthew Edwards}, \bibinfo{person}{Pauline
  Anthonysamy}, \bibinfo{person}{Sylvain Frey}, {and} \bibinfo{person}{Awais
  Rashid}.} \bibinfo{year}{2020}\natexlab{}.
\newblock \showarticletitle{"{{So}} If {{Mr Blue Head}} Here Clicks the Link.."
  {{Risk Thinking}} in {{Cyber Security Decision Making}}}.
\newblock \bibinfo{journal}{\emph{ACM Transactions on Privacy and Security}}
  \bibinfo{volume}{24}, \bibinfo{number}{1} (\bibinfo{date}{Nov.}
  \bibinfo{year}{2020}), \bibinfo{pages}{5:1--5:29}.
\newblock
\showISSN{2471-2566}
\urldef\tempurl%
\url{https://doi.org/10.1145/3419101}
\showDOI{\tempurl}


\bibitem[\protect\citeauthoryear{Stevens and Norman}{Stevens and
  Norman}{2016}]%
        {stevensIndustryExpectationsSoft2016}
\bibfield{author}{\bibinfo{person}{Matt Stevens} {and} \bibinfo{person}{Richard
  Norman}.} \bibinfo{year}{2016}\natexlab{}.
\newblock \showarticletitle{Industry Expectations of Soft Skills in {{IT}}
  Graduates: A Regional Survey}. In \bibinfo{booktitle}{\emph{Proceedings of
  the {{Australasian Computer Science Week Multiconference}}}}
  \emph{(\bibinfo{series}{{{ACSW}} '16})}. \bibinfo{publisher}{{Association for
  Computing Machinery}}, \bibinfo{address}{{New York, NY, USA}},
  \bibinfo{pages}{1--9}.
\newblock
\showISBNx{978-1-4503-4042-7}
\urldef\tempurl%
\url{https://doi.org/10.1145/2843043.2843068}
\showDOI{\tempurl}


\bibitem[\protect\citeauthoryear{{Taylor-Jackson}, McAlaney, Foster, Bello,
  Maurushat, and Dale}{{Taylor-Jackson} et~al\mbox{.}}{2020}]%
        {taylor-jacksonIncorporatingPsychologyCyber2020}
\bibfield{author}{\bibinfo{person}{Jacqui {Taylor-Jackson}},
  \bibinfo{person}{John McAlaney}, \bibinfo{person}{Jeffrey~L. Foster},
  \bibinfo{person}{Abubakar Bello}, \bibinfo{person}{Alana Maurushat}, {and}
  \bibinfo{person}{John Dale}.} \bibinfo{year}{2020}\natexlab{}.
\newblock \showarticletitle{Incorporating {{Psychology}} into {{Cyber Security
  Education}}: {{A Pedagogical Approach}}}. In
  \bibinfo{booktitle}{\emph{Financial {{Cryptography}} and {{Data Security}}}}
  \emph{(\bibinfo{series}{Lecture {{Notes}} in {{Computer Science}}})},
  \bibfield{editor}{\bibinfo{person}{Matthew Bernhard}, \bibinfo{person}{Andrea
  Bracciali}, \bibinfo{person}{L.~Jean Camp}, \bibinfo{person}{Shin'ichiro
  Matsuo}, \bibinfo{person}{Alana Maurushat}, \bibinfo{person}{Peter~B.
  R{\o}nne}, {and} \bibinfo{person}{Massimiliano Sala}} (Eds.).
  \bibinfo{publisher}{{Springer International Publishing}},
  \bibinfo{address}{{Cham}}, \bibinfo{pages}{207--217}.
\newblock
\showISBNx{978-3-030-54455-3}
\urldef\tempurl%
\url{https://doi.org/10.1007/978-3-030-54455-3_15}
\showDOI{\tempurl}


\bibitem[\protect\citeauthoryear{Thomson and Oppenheimer}{Thomson and
  Oppenheimer}{2016}]%
        {thomsonInvestigatingAlternateForm2016}
\bibfield{author}{\bibinfo{person}{Keela~S Thomson} {and}
  \bibinfo{person}{Daniel~M Oppenheimer}.} \bibinfo{year}{2016}\natexlab{}.
\newblock \showarticletitle{Investigating an Alternate Form of the Cognitive
  Reflection Test}.
\newblock \bibinfo{journal}{\emph{Judgment and Decision making}}
  \bibinfo{volume}{11}, \bibinfo{number}{1} (\bibinfo{year}{2016}),
  \bibinfo{pages}{99}.
\newblock
\showISSN{1930-2975}


\bibitem[\protect\citeauthoryear{Towse, Levine, Petre, Bandara, Lopez, Rashid,
  Rauf, Sharp, Tun, {van der Linden}, and Nuseibeh}{Towse
  et~al\mbox{.}}{ress}]%
        {towseCaseUnderstandingSecure2020}
\bibfield{author}{\bibinfo{person}{John Towse}, \bibinfo{person}{Mark Levine},
  \bibinfo{person}{Marian Petre}, \bibinfo{person}{Arosha Bandara},
  \bibinfo{person}{Tamara Lopez}, \bibinfo{person}{Awais Rashid},
  \bibinfo{person}{Irum Rauf}, \bibinfo{person}{Helen Sharp},
  \bibinfo{person}{Thein Tun}, \bibinfo{person}{Dirk {van der Linden}}, {and}
  \bibinfo{person}{Bashar Nuseibeh}.} \bibinfo{year}{2020 (in
  press)}\natexlab{}.
\newblock \showarticletitle{The {{Case}} for {{Understanding Secure Coding}} as
  a {{Psychological Enterprise}}}.
\newblock \bibinfo{journal}{\emph{Cyberpsychology, Behavior, and Social
  Networking}} (\bibinfo{year}{2020 (in press)}).
\newblock


\bibitem[\protect\citeauthoryear{Wurster and {van Oorschot}}{Wurster and {van
  Oorschot}}{2008}]%
        {wursterDeveloperEnemy2008}
\bibfield{author}{\bibinfo{person}{Glenn Wurster} {and} \bibinfo{person}{Paul
  {van Oorschot}}.} \bibinfo{year}{2008}\natexlab{}.
\newblock \showarticletitle{The Developer Is the Enemy}. In
  \bibinfo{booktitle}{\emph{Proceedings of the 2008 {{New Security Paradigms
  Workshop}}}} \emph{(\bibinfo{series}{{{NSPW}} '08})}.
  \bibinfo{publisher}{{Association for Computing Machinery}},
  \bibinfo{address}{{New York, NY, USA}}, \bibinfo{pages}{89--97}.
\newblock
\showISBNx{978-1-60558-341-9}
\urldef\tempurl%
\url{https://doi.org/10.1145/1595676.1595691}
\showDOI{\tempurl}


\end{thebibliography}

\end{document}